\documentclass{article}

\usepackage[english]{babel}

\usepackage{upgreek}
\newcommand{\kbt}{k_\mathrm{B}T}

\usepackage[letterpaper,top=2cm,bottom=2cm,left=3cm,right=3cm,marginparwidth=1.75cm]{geometry}

\usepackage{amsmath}
\usepackage{graphicx}
\usepackage[colorlinks=true, allcolors=blue]{hyperref}

\title{Early stage of Erythrocyte Sedimentation Rate test: Fracture of a high-volume-fraction gel}
\author{Thomas John, Lars Kaestner, Christian Wagner, Alexis Darras\\{\textit Experimental Physics, Saarland University, 66123 Saarbruecken, Germany}}

\begin{document}
\maketitle

\begin{abstract}
Erythrocyte Sedimentation Rate (ESR) is a clinical parameter used as a non-specific marker for inflammation, and recent studies have shown that it is linked to the collapse of the gel formed by red blood cells (RBCs) at physiological hematocrits (i.e. RBC volume fraction). Previous research has suggested that the delay time before the sedimentation process is related to the formation of fractures in the gel. Moreover, RBC gels present specific properties due to the anisotropic shape and flexibility of the RBCs. Namely, the onset of the collapse is reached earlier and the settling velocity of the gel increases with increasing attraction between the RBCs, while gel of spherical particles show the opposite trend. Here, we report experimental observations of the gel structure during this onset and suggest an equation modeling this initial process as fracturing of the gel. We demonstrate that this equation provides a model for the motion of the interface between blood plasma and the RBC gel, along the whole time span. We also observe that the increase in the attraction between the RBCs modifies the density of fractures in the gel, which explains why the gel displays a decrease in delay time when the aggregation energy between the RBCs increases. Our work uncovers the detailed physical mechanism underlying the ESR and provides insights into the fracture dynamics of a RBC gel. These results can improve the accuracy of clinical measurements.
\end{abstract}

\section{Introduction}

The Erythrocyte Sedimentation Rate (ESR) is a blood test that measures how quickly red blood cells settle in a test tube, and has been used for centuries to diagnose and monitor inflammatory diseases \cite{kushner1988acute,grzybowski2011edmund,bedell1985,tishkowski2020erythrocyte,passos2021erythrocyte}. It is a non-specific test that is sensitive to increases in fibrinogen and other plasma components \cite{brigden1999clinical,greidanus2007use,gray1942}. Recent research has shown that it may also be useful in detecting abnormally-shaped red blood cells \cite{darras2021acanthocyte,rabe2021erythrocyte,peikert2022xk}. Despite its widespread use, the physical mechanisms governing the ESR are not yet fully understood.
It has recently been demonstrated that the cause of this sedimentation is the gravitational collapse of the percolating network, also known as gel, formed by the RBCs \cite{darras2022,dasanna2022erythrocyte}. Similarly to colloidal gels, this collapse presents an initial delay time, during which no or negligible sedimentation is observed \cite{hung1994erythrocyte,teece2014gels,buscall2009towards,gopalakrishnan2006linking,padmanabhan2018gravitational,bartlett2012sudden,derec2003rapid,poon1999delayed,allain1995aggregation,harich2016}. The origin of colloidal gel sedimentation delay is still debated, however it is likely to be associated with gel aging and the development of cracks for fluid flow within the gel \cite{teece2014gels,buscall2009towards,gopalakrishnan2006linking,padmanabhan2018gravitational,bartlett2012sudden,pribush2010mechanism1,lindstrom2012}. Surprisingly, contrary to colloidal hard spheres suspensions, an increase in attractive interactions between RBCs results in gel destabilization, leading to faster structure rearrangement and apparition of cell-depleted cracks, which collapses faster \cite{dasanna2022erythrocyte,darras2022imaging}. This feature likely contributed to the establishment of the ESR as a medical tool, as a shorter delay time and an increased collapse velocity are additive for the typical medical read-out, which considers the average velocity of the interface during the first hour \cite{bedell1985,brigden1999clinical,tishkowski2020erythrocyte}.
In this study, we conducted experiments at different length scales to investigate the dominant mechanism of the fracture process in RBC gels, and compare it to a theoretical model from prior literature \cite{varga2018}. We demonstrated that higher RBC aggregation energy results in more fractures in the gel. Moreover, we derived a new equation for the delay part of our previous model for the macroscopic interface velocity \cite{darras2022,dasanna2022erythrocyte}. These fundamental findings can be used to extract more rigorous and reproducible parameters from erythrocyte sedimentation rate measurements in clinical context \cite{darras2023PhyDriv}.

\section{Microscopic experiments}

\subsection{Microscopic scale observations of the fracture}

We performed experiments using light sheet microscopy (Z1, Zeiss, Jena, Germany) as described in a previous methodological publication \cite{darras2022imaging}. This technique allows enough resolution to extract the velocity field of the RBCs in the obtained image sequences using Particle Image Velocimetry (PIV, through PIVLab \cite{thielicke2021particle}), but only probes a small part of the gel close to its lateral edge. More accurately, only a depth of approximately $100~\mathrm{\upmu m}$ on an area around $1~\mathrm{mm^2}$ could be observed, while the whole cylindric sample has a height around 3~cm and a diameter of 1.6~mm. The area where the PIV could reliably be performed is even smaller, since absorption and diffraction of the laser light decrease the overall intensity of the picture around 700~$\mathrm{\upmu}$m from the border of the sample.

\begin{figure*}[htbp]
 \centering
 \includegraphics[width=0.9\linewidth]{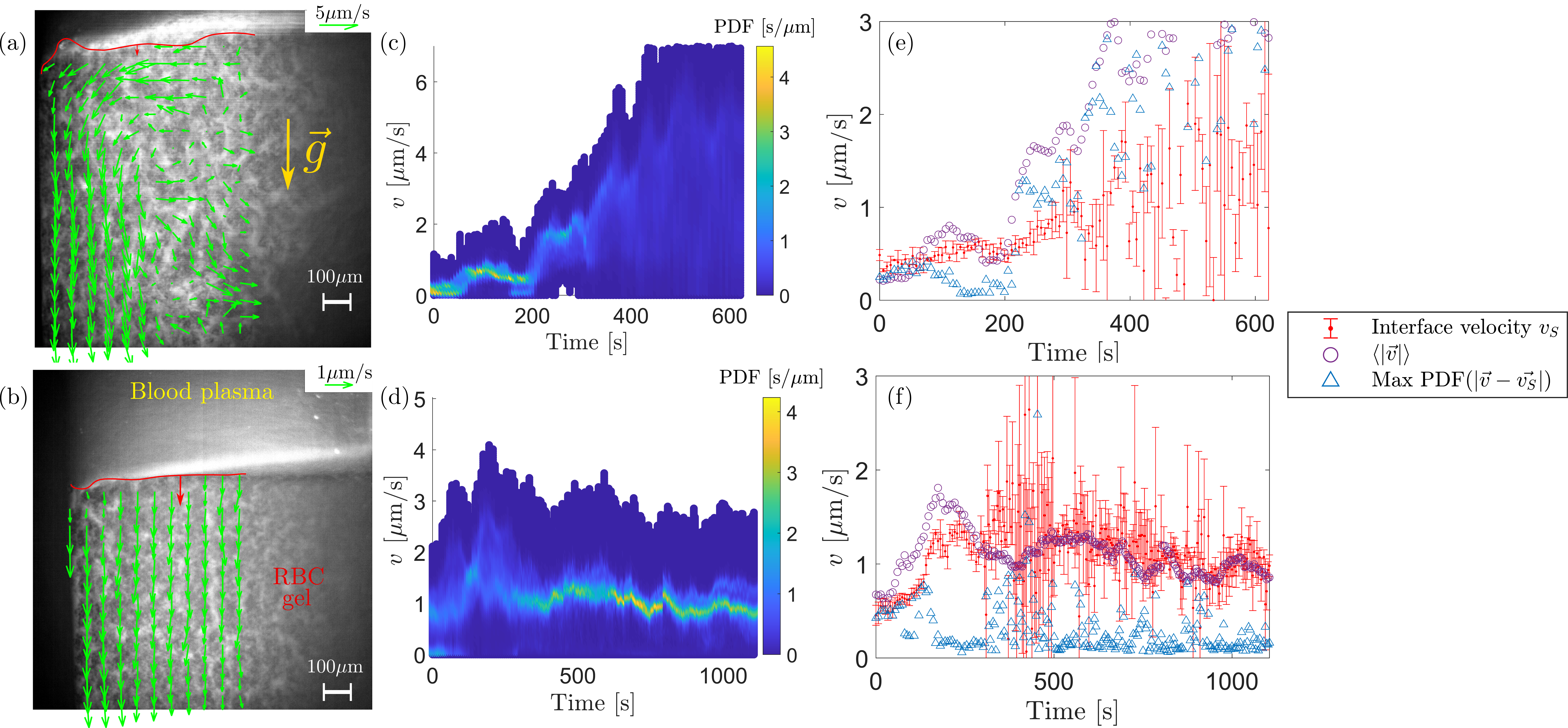}
 \caption{\textbf{Flow Field Characterization using Light Sheet Microscopy.} Two representative identically repeated experiments (first and second line) are shown with samples from the same RBC suspension, locally showing qualitatively different velocity fields. $(a,b)$ Reconstructed images, with sampled velocity field at the end of the experiments (resp. $\approx 10$ and $20$ min). See Movies S1 and S2 for the whole characterization. $(c,d)$ Evolution of the Probability Density Function (PDF) of the velocity modulus over time. $(e,f)$ Comparison of the velocity of the interface with the velocity of the flow field and its average modulus. In the upper row, the experiment displays a complex velocity field (panel $(a)$) due to a partial gel fluidization, which propagates from the right of the image (see Movie S1). The corresponding PDF (panel $(c)$) starts with a peak that gradually disappears as the network loses its integrity, while the average velocity (panel $(e)$) increases above the velocity of the network interface. In the lower row, the experiment exhibits a cohesive downward velocity with no significant fluidization ($(b)$, Movie S2). Instead, the gel reaches a cohesive downward velocity. The PDF $(d)$ shows a peak that persists over time and closely follows the interface velocity $(f)$. The maximal interface velocity in both experiments ($(e)$ and $(f)$) is similar (around $1.3~\mathrm{\upmu m/s}$), consistent with the maximum macroscopic interface velocity for similar samples (hematocrit $\phi=0.45$)\cite{darras2022}.}
 \label{Fig1}
\end{figure*}

As illustrated in Fig.\ref{Fig1}, and displayed in Supplementary Movies S1 and S2, when repeating experiments with samples from the same suspension, we obtained qualitatively different behaviors of the velocity field, even though the global interface velocity is reproducible. Specifically, we noted instances where the gel fluidized within the field of view, while in other cases, it exhibited a cohesive behavior, resembling a solid translation.
However, we also extracted a more global parameter by extracting the velocity of the interface of the RBC gel. In order to accomplish this, we detect the position of the interface by locating the height with the strongest vertical intensity falling edge. This is achieved by averaging the vertical intensity over a horizontal width of 250~$\mathrm{\upmu m}$ at each point to ensure both robust and accurate detection of the interface position.
The average velocity reached by the interface is reproducible within experimental accuracy, which implies that the macroscopic dynamics of the entire sample is reproducible. Notably, velocities significantly higher than the interface velocity are observed within the gel when the velocity field is not homogeneous (Fig. \ref{Fig1}(a-e)). However, when a nearly solid translation of the RBC gel is observed in the field of view (Fig. \ref{Fig1}(b-f)), the peak velocity observed in the velocity probability density function matches the interface velocity.  This suggests that the RBC gel undergoes partial fluidization from an initial streamer, but this fluidization does not spread throughout the entire sample. Consequently, the rest of the sample follows the overall compaction of the structure, which ultimately determines the surface velocity.
In order to confirm and generalize these conclusions, we also performed observations at bigger scales, as described in the next sections.

\subsection{Mesoscopic scale observation of the fracture}

\begin{figure*}
 \centering
 \includegraphics[width=0.85\linewidth]{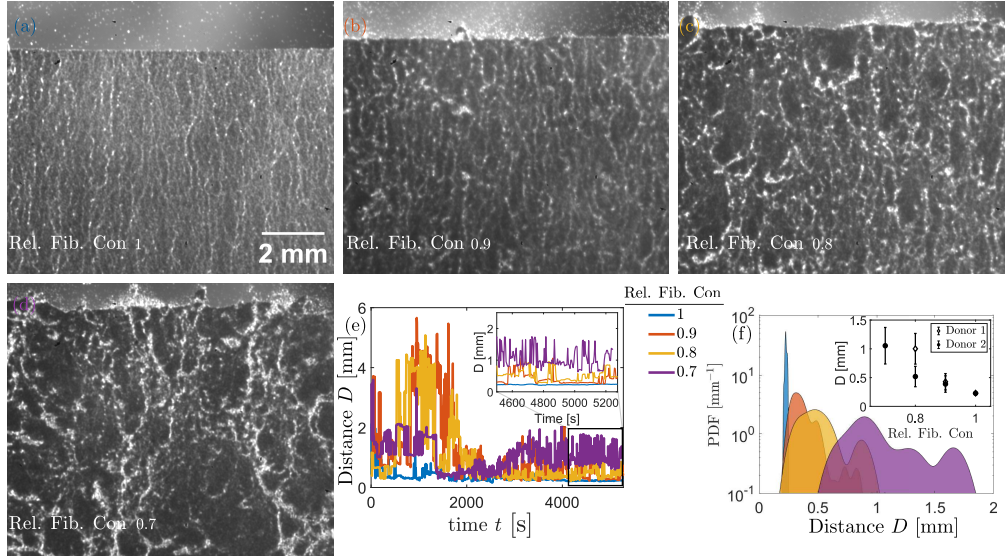}
 \caption{\textbf{Mesoscopic structures revealed by Infrared light transmission through thin samples.} $(a-d)$ Images captured after apparition of the streamers (i.e. after approx. 1$\,$h 15$\,$min), showcasing samples with decreasing concentrations of fibrinogen from $(a)$ to $(d)$. Corresponding movies of the full experiments $(a)$ to $(d)$ are provided as Supplementary Movies S3 to S6, respectively. $(e)$ Time evolution of the characteristic distance between the streamers, computed as the position of the first non-zero maxima of the horizontal auto-correlation of the images. $(f)$ PDF of the first non-zero peak of the horizontal auto-correlation function, after $4500\,$s (i.e. the data shown as inset in panel $(e)$). Inset shows averaged distance between the streamers plotted against the relative concentration of fibrinogen (Rel. Fib. Con.), for two different donors. As the fibrinogen concentration increases, the characteristic distance between the streamers decreases.}
 \label{Fig2}
\end{figure*}
 
 To investigate the larger-scale structure of RBC gels, we utilized microscopy with infrared light transmission through thin samples. We followed a similar procedure outlined in a previous paper which used blue light, \cite{darras2022imaging} however, we replaced the blue LED source with a halogen lamp (Nikon, LHS-H100P-1). The emitted light was filtered by an infrared long pass filter with a cutting wavelength of 950~$\mathrm{\upmu m}$ (Neewer, IR950). Using infrared light provided us with higher transmission through RBCs, revealing greater detail in the structure than the blue light, which was less sensitive to the thickness of the sample. 
 We conducted experiments with an adjusted hematocrit of $\phi=0.45$ and various dilutions of autologous plasma with serum. Serum can be considered as plasma without fibrinogen, as the coagulation cascade occurs prior to serum extraction. This method allowed us to dilute the fibrinogen content while retaining other plasma proteins, which effectively tunes the attractive forces between RBCS \cite{issaq2007serum,dasanna2022erythrocyte,darras2023PhyDriv,brust2014}. We observed significant differences in the RBC gel structure for relative concentration of plasma in the liquid phase varying from $0.7$ (e.g. 0.7~mL of plasma are mixed with 0.3~mL of serum) to $1$ (RBCs are suspended in pure plasma).
We were able to image an area of almost 1~cm$\times$0.87~cm  in the samples, with a total width $\times$ height $\times$ thickness of approximately  1.7~cm$\times$7~cm$\times$150~$\mathrm{\upmu m}$. As shown in Fig.\ref{Fig2}$(a-d)$ and Supplementary Movies S3 to S6, this setup revealed that during the initial stages of the sedimentation process, the RBC gel fractures, resulting in vertically oriented streamers at various horizontal intervals. 
 To quantify the evolution of the characteristic distance $D$ between the streamers, we computed the position of the first non-zero maxima of the horizontal auto-correlation of the image. Those quantity shows strong fluctuation, however after a transition time of approximately 4500~s it shows a clear decreasing trend for all fibrinogen concentrations, see Fig. \ref{Fig2}$(e)$.  With  increasing fibrinogen concentrations, i.e. stronger RBC interactions, the distance $D$ between the streamers decreased significantly, as shown in Fig.\ref{Fig2}$(f)$.

In summary, our experimental observations demonstrated that the RBC gel is locally fluidized into streamers at the initial stage of its sedimentation, in a process similar to the observations reported in simulations of the onset of colloidal gel settling \cite{varga2018}. However, the fluidization of the structure never occurs over the whole sample. Eventually, the network of streamers stabilizes, i.e. the streamers stop forming or growing, and the RBC network undergoes a smoother reorganization, which can be described as the compression of a porous material \cite{darras2022}. 

\subsection{Macroscopic scale measurements}

At larger scales, one observes a motion of a sharp interface between cell-free plasma and sedimenting RBCs. The average velocity of this interface over the first hour is actually the parameter measured to perform an ESR test. This measurement is typically done by assessing the position of the interface at the beginning of the experiments and after one hour of leaving the sample at rest. To complete our observations with macroscopic data, we conducted experiments similar to those in a previous study \cite{dasanna2022erythrocyte}, where we manipulated the concentration of fibrinogen by mixing serum and plasma for suspensions of RBCs, with the hematocrit held constant at $\phi=0.45$. The experimental setup is depicted in Figure \ref{Fig3}$(a)$, with image analysis data presented in Figure \ref{Fig3}$(b,c)$. For a further detailed explanation of the picture post-processing, please refer to our earlier methodological publication \cite{darras2023PhyDriv}; The velocity points in Fig.\ref{Fig3}$(b)$ are extracted as follows. We first compute the finite difference between two successive height measurements (Fig.\ref{Fig3}$(c)$), divided by the time interval of 60\,s between two consecutive pictures. The resulting values are fitted by a smoothing spline. The smoothing spline is then evaluated at the time of the images to obtain the open circles in Fig.\ref{Fig3}$(b)$.

\section{Modeling equations and comparison to macroscopic measurements}

\subsection{Equations of interface motion}

\begin{figure*}[t!]
 \centering
 \includegraphics[width=0.8\linewidth]{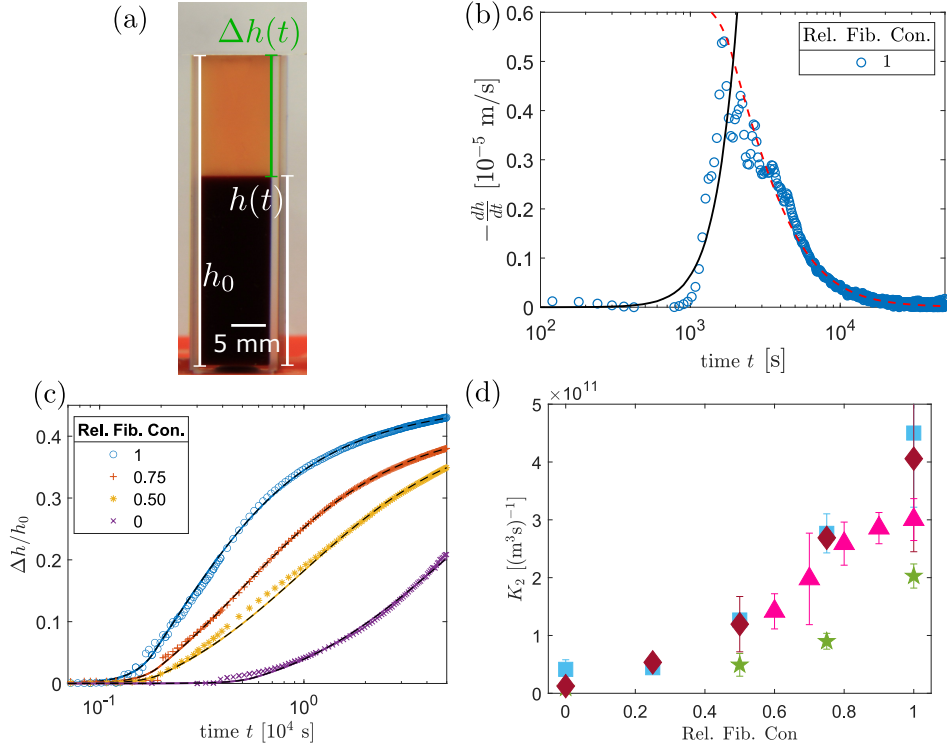}
 \caption{\textbf{Macroscopic measurements of Erythrocyte Sedimentation Rate.}$(a)$ Image of a cuvette used for macroscopic experiments, annotated with the measured heights. $(b)$ Velocity of the interface along time for a representative experiment. Points are data gathered from experiments, while the black curve is the first line of Eq.(\ref{EqFitGlob}) (i.e. the prediction of Eqs.(\ref{dRdt},\ref{MeanUp}) combined) and the red dashed curve is the second line. Values of the fit parameters for Eq.(\ref{EqFitGlob}) are obtained by fitting the position of the interface (panel $(c)$). $(c)$ Evolution of the RBC gel height along time. The points are measurements from image analysis, and the curves depict fits from Eq.(\ref{EqFitGlob}). $(d)$ Evolution of the $K_2$ parameter from Eq.(\ref{EqFitGlob}), as a function of the relative fibrinogen concentration. Different symbols are associated to different donors. As expected from the decreasing trend observed in Fig.\ref{Fig2}$(f)$ and $K_2\propto D^{-2}$, increasing fibrinogen concentration results in an increase in $K_2$.}
 \label{Fig3}
\end{figure*}

In a previous manuscript by Varga \textit{et al.} \cite{varga2018}, the instability that causes the formation and spread of a fluidized portion of a settling gel, known as a 'streamer', was investigated both analytically and numerically. However, their model contains several geometrical constants and parameters that are difficult to estimate experimentally. In this paper, we present an approach for simplifying their model by using only the leading terms to derive a system of differential equations that can be fitted to experimental data with just two fit parameters.


To ensure clarity, we will begin by defining the key parameters of the gel. The red blood cells (RBCs) in the gel have a characteristic radius of $a\approx4\,\mathrm{\upmu m}$ and experience surface attraction that is described by a potential well with a depth of $U$of (10 to 20)$\,\kbt$ and a width $\Delta$ between (10 to 100)$\,\mathrm{nm}$ \cite{brust2014,liu2006rheology,neu2002depletion,baskurt2011red,lee2016optical,linss1991thick}. The surrounding fluid has a viscosity of $\eta\approx10^{-3}\,\mathrm{Pa\,s}$ \cite{kesmarky2008plasma}, and both the RBCs and fluid molecules have thermal energy $\kbt$. However, the density difference $\rho$ between the cells and the fluid is 10 to $100\,\mathrm{kg/m^3}$ \cite{trudnowski1974specific}. Additionally, we will refer to the geometrical constants $d_i$ introduced by Varga \textit{et al.} \cite{varga2018} in their model.

The growth of the streamer radius $R$ over time $t$ is determined by Eq. (2.10) in Varga \textit{et al.} \cite{varga2018}:
\begin{align}
    \frac{dR}{dt}&=\frac{K}{\phi^{1/3}} R^{-1/3} e^{R/R^*},\label{dRdt}\\
    \text{where~} K&=\frac{2d_1d_2}{9}\frac{a^{1/3}Ue^{-U/(d_3\kbt)}}{\eta\Delta^2}\label{Kdef}\\
    \text{and~} R^*&=\frac{2}{d_4}\frac{\kbt}{\rho g a^2 \Delta}.\label{Rsdef}
\end{align}

From its definition, $R^*$ can be described as an effective gravitational length. Assuming that all dimensionless geometrical constants $d_i$ are of the order of unity, we can estimate $K\in\left[10^{-11}\,\text{to}\,10^{-7}\right]\,\mathrm{m^{4/3}/s}$ and $R^*\in\left[10^{-5}\,\text{to}\,10^{-4}\right]\,\mathrm{m}$.

We assume that the plasma flows mainly through the streamer, i.e. that the permeability $\kappa$ of the undisturbed gel is negligible. This assumption is consistent with the fact that the initial velocity of the gel interface is below the experimental resolution. We also disregard the inverse Navier's slip length, $\lambda$, at the border of the streamers. Under these assumptions, Eq. (2.13) of Varga \textit{et al.}\cite{varga2018} determines the average upward plasma velocity, $\langle u_f\rangle$, as $\langle u_f\rangle=K_2 R^4$, with $K_2=\frac{\pi}{8}\frac{\rho g}{\eta D^2}$, where the distance $D$ is assumed by Varga \textit{et al.} to be equal to the cross-sectional length of the sample $L$. Note that this expression of $K_2$ is valid only if there is a single streamer, as observed in numerical simulations performed on a spatially limited system. However, we experimentally observed several streamers, as illustrated in Fig. \ref{Fig2}$(a-d)$. This implies that we should rather consider $D$ to be the characteristic distance between two neighboring streamers at the interface of the gel, i.e. $D^2=L^2/N$, with $N$ the number of streamers which are distributed all over the cross-section area of the sample.
Assuming volume conservation, the velocity of the interface can then be expressed as
\begin{equation}
\frac{dh}{dt}=-\langle u_f\rangle \frac{(1-\phi)}{\phi}=-K_2 R^4 \frac{(1-\phi)}{\phi},
\label{MeanUp}
\end{equation}
where the average volume fraction $\phi$ of the gel is dependent on both the initial volume fraction $\phi_0$ and height $h_0$, due to volume conservation of the RBCs. Specifically, $\phi$ can be expressed as $\phi=\phi_0\,{h_0}/{h}$.

Experimentally, we observed that the growth of streamer radius $R$ is limited over time. Intermediate-scale experiments (Fig. \ref{Fig2}, Movies S3-6) indeed showed that the diameters of the streamers saturate, and their positions stabilize. The discrepancy could be attributed to one of the assumptions made in Varga \textit{et al.}'s model \cite{varga2018}, which states that the volume fraction of particles inside the streamer is constantly similar to the bulk volume fraction. This approximation is explicitly mentioned when they assess the number of particles in a streamer, and implicitly used when they assessed the value of the flux of particles per unit area into and out of the open streamer $j_{in}$ and $j_{out}$ (Eqs. (2.7) and (2.8) in \cite{varga2018}). However, both their numerical simulations and our experiments demonstrate that the volume fraction within the streamers decreases. Since both fluxes are proportional to $3\phi/(4\pi a^3)$ , if the volume fraction $\phi$ inside the streamers decreases significantly, these fluxes should therefore vanish, which explains that the streamers stabilize and the gel interface reaches a maximal velocity.

As shown by Darras \textit{et al.} \cite{darras2022}, once the maximal velocity and underlying structure of the gel are achieved, the collapse of the RBC gel can be modeled as a porous medium compressing under its own weight. The velocity of the interface is described as $dh/dt=-\frac{\rho g a^2}{\gamma \eta}\frac{(\phi_m-\phi)^3}{\phi(1-\phi)}$, where $\gamma$ is a dimensionless characteristic time of the system, and $\phi_m$ is the maximal volume fraction reached by the RBCs in the gel at the end of the sedimentation process. In summary, the interface velocity can be expressed as
\begin{eqnarray}
-\frac{dh}{dt}=\min  
    \begin{cases}
    K_2 R(t)^4 \frac{(1-\phi)}{\phi}\\
    \frac{\rho g a^2}{\gamma \eta}\frac{(\phi_m-\phi)^3}{\phi(1-\phi)}&
    \end{cases},
\label{EqFitGlob}
\end{eqnarray}
with $R(t)$ from Eq. (\ref{dRdt}). This equation can be fitted to macroscopic experimental data, using in total four fit parameters $K$, $K_2$, $\gamma$ and $\phi_m$.

\subsection{Model fitting}

To numerically solve Eq.(\ref{EqFitGlob}), an initial value $R(t=0)$ is required. However, the choice of this value has little effect on the time when $dR/dt$ diverges as long as $R_0\ll R^*$. This is because analytical results from Varga \textit{et al.} also predict a finite time divergence of $dR/dt$ for $R(0)=0$ (see their Eq.(2.15) and Fig. 7 in \cite{varga2018}). Therefore, we used $R(0)=1\mathrm{\upmu m}$, which is the same order of magnitude as the holes observed in 2D percolating networks of RBCs \cite{dasanna2022erythrocyte}. We used $R^*$ as a fit parameter, initially constrained in the range of $\left[10^{-5}\,\text{to}\,10^{-3}\right]\,\mathrm{m}$ based on estimations from $d_4=1$ and $\rho\in\left[10\,\text{to}\,100\right]\,\mathrm{kg/m^3}$, which led to $R^*\in[5.\times10^{-6}\,\text{to}\,5.\times10^{-4}]\,\mathrm{m}$. However, the choice of $R^*$ in this interval did not have a significant influence on the sum of the square residuals of the initial fits we tried. The obtained values of $R^*$ all fell within the range of $(3.0\pm0.1)\,10^{-4}\,\mathrm{m}$. Therefore, we used $R^*=3.10^{-4},\mathrm{m}$ as a fixed parameter for all fits. 
As the Eq.(\ref{EqFitGlob}) is highly non-linear, the success of the fit convergence depends on the initial guess of the other fit parameters. To simplify this problem, we first estimated the values of $\phi_m$ and $\gamma$ using the previous approximation that $h(t)=h_0$ if $t<t_0$ \cite{dasanna2022erythrocyte}. These estimated values of $\phi_m$ and $\gamma$ were then used as initial guesses for the fit of Eq.(\ref{EqFitGlob}). Additionally, we set the initial value of $K_2$ such that both time derivatives in Eq.(\ref{EqFitGlob}) were equal at $R=50\,\mathrm{\upmu m}$. This $R=50\,\mathrm{\upmu m}$ roughly corresponds to the radius of the depleted areas observed for the full plasma sample (Fig.\ref{Fig2}$(a)$, Movie S3). The parameter $K$ is calculated through its definition in Eq.(\ref{dRdt}) with $d_1=d_2=1$, $\Delta=10^{-8}\,\mathrm{m}$, and $U=15\kbt$, which yielded $K\approx10^{-9}\,\mathrm{m^{3/4}/s}$ as the initial guess.
The fits obtained using this protocol are in good agreement with measured data, as shown in Fig. \ref{Fig3}$(b,c)$. The parameter $K_2$ exhibits a significant trend as a function of the fibrinogen concentration, as illustrated in Fig. \ref{Fig3}$(d)$, while the values of $K$ are almost constant within the range $(7\pm2)\,10^{-10}\,\mathrm{m^{4/3}/s}$ (see Supp. Fig. S1). The behaviors of $\gamma$ and $\phi_m$ are consistent with the trends reported in \cite{dasanna2022erythrocyte}, see also Supp. Fig. S1. As expected, due to the decrease in interdistance $D$ between the streamers as the concentration of fibrinogen increases, we observe that $K_2\propto D^{-2}$ increases with increasing fibrinogen concentration. This change in the geometry of the gel structure is probably related to the change in pore size that static 2D networks of RBCs exhibit when their interaction energy is modified \cite{dasanna2022erythrocyte}. Indeed, a higher amount of bigger pores in the initial network implies more probable seeding points for the fracture of the gel. The porosity of the collapsing gel therefore increase with an increase in the fibrinogen concentration.




\section{Implications for clinical ESR tests}

In current medical applications, only a two time-points measurement is considered to estimate the average sedimentation velocity during the first hour of the sedimentation process. Some of the current features of the standard ESR measurement are that there is no lower bound for the normal range, and no correction as a function of the sample hematocrit is universally recognized \cite{kratz2017icsh,kim2004comparative,dintenfass1974erythrocyte,borawski2001hematocrit}. 
Previous clinical studies have already highlighted that the lack of lower bound is related to the fact that the maximum velocity is often reached after the first hour, where the measurement is done \cite{hung1994erythrocyte,woodland1996erythrocyte,holley1999influence}. Since both time derivatives in Eq.(\ref{EqFitGlob}) have a different dependency with $\phi$, it is now clear from our model that a simple scaling of the one time-point measurement can not be rigorously obtained. However, if one records regularly the position of the interface over a longer period of time (approx. 2h, as already suggested in some protocols \cite{darras2021acanthocyte,rabe2021erythrocyte,hung1994erythrocyte,woodland1996erythrocyte}), as is now easily enabled by automation, one can extract the maximum velocity $\left|dh/dt\right|$, which scales as $\left|dh/dt\right|\propto \frac{(\phi_m-\phi)^3}{\phi(1-\phi)}$ \cite{darras2022}. This more detailed analysis protocol could therefore provide a lower bound for the normal range of ESR, which could be used as a clinical tool to detect rare diseases, such as neuroacanthocytosis syndromes, which presents a significantly slower ESR \cite{rabe2021erythrocyte,darras2021acanthocyte}.

\section{Conclusions}
The experiments performed at various length scales have revealed that the initial collapse of the RBC gel is initiated by local instabilities that lead to the appearance of multiple streamers. The spatial distribution of these streamers depends on the interaction energy between the cells. This is consistent with earlier observations that higher aggregation between RBCs results in larger pore sizes, thereby increasing the possible seeding points for the emergence of streamers within the bulk. The increase in the average distance between the streamers qualitatively explains the macroscopic characteristics of the collapse: with higher cell aggregation, more streamers appear and the gel collapses sooner. On a fundamental level, these results lead to a continuous model for the gravitational collapse of a gel with a delay time. We have successfully connected the microscopic rearrangements of the gel structure to the macroscopic velocity of the gel interface. 
It is worth noting that the physical peculiarities of RBC aggregates are crucial for the aforementioned mechanisms. Indeed, the increase in delay time with higher aggregation energy is due to the geometry of the RBC aggregates, which therefore contributes to making them a unique suspension. Understanding these peculiarities is of immense significance, particularly in clinical contexts. The present results provide a systematic approach to extract detailed parameters of the erythrocyte sedimentation rate, which can lead to a more rigorous and precise description of the sedimentation velocity than the current clinical standard, which only considers the average sedimentation velocity during the first hour by a two time-points measurement.

\section{Material and methods}

Blood sample collection and experiments were approved by the “\"Arztekammer des Saarlandes”, ethics votum 51/18, and performed after informed consent was obtained according to the Declaration of Helsinki. Blood was collected in standard EDTA-anticoagulated blood, as well as standard serum tubes.

\section{Acknowledgments} 
This work was supported by the research unit FOR 2688 - Wa1336/12 of the German Research Foundation, and by the Marie Skłodowska-Curie grant agreement No. 860436—EVIDENCE. A.D. acknowledges funding by the Young Investigator Grant of the Saarland University. The authors gratefully acknowledge Prof. Paulo E. Arratia (University of Pennsylvania) for fruitful discussions on the manuscript.

\bibliography{pnas-sample.bib}
\bibliographystyle{abbrv}

\end{document}